\documentclass{article}

\usepackage[utf8]{inputenc}
\usepackage[T2A]{fontenc}

\usepackage[english]{babel}
\usepackage[tbtags]{amsmath}
\usepackage{amsfonts,amssymb,mathrsfs,amscd,comment}
\usepackage{umnbib} % для формления списка литературы в подбор

\usepackage{tikz-cd}        % commutative diagram

\newtheorem{thm}{Theorem}

\newtheorem{lemma}[thm]{Lemma}

\newcommand{\cA}{\mathcal A}
\newcommand{\fA}{\mathfrak A}

\newcommand{\cI}{\mathcal I}
\newcommand{\cJ}{\mathcal J}

\newcommand{\cL}{\mathcal L}

\newcommand{\cS}{\mathcal S}

\newcommand{\bbC}{\mathbb C}

\newcommand{\bbN}{\mathbb N}

\newcommand{\bbZ}{\mathbb Z}

\newcommand{\bF}{\mathbf F}
\newcommand{\bu}{\mathbf u}

\newcommand{\bw}{ w}

\newif\ifprivate
\privatefalse

 \numberwithin{equation}{section}

\def\???{\ifprivate {\bf {???}} \marginpar{{\Huge {\bf ?}}}\else \fi}
\numberwithin{equation}{section}

\renewcommand{\theequation}{\arabic{equation}}

\begin{document}

\title{\bf\large\MakeUppercase{Finite dimensional reductions of integrable  differential-difference equations
}}

\author{A.V.~Mikhailov}
\date{}

\maketitle

\makeatletter
\renewcommand{\@makefnmark}{}
\makeatother

Integrable partial differential equations, such as the Korteweg--De Vries (KdV) equation, admit infinite hierarchies of commuting higher symmetries. Their symmetry reductions give rise to integrable finite-dimensional dynamical systems solvable in terms of Abelian functions. This underlies the finite-gap integration method for the KdV equation, introduced by S. P. Novikov \cite{Novikov1974} and subsequently extended to many other integrable systems. In this paper, we introduce a new and more general class of reductions for integrable differential-difference equations, yielding integrable finite-dimensional systems in both commutative and noncommutative settings. We use the Volterra hierarchy for illustrations.

{\bf 1.} Let $\cA=(\bbC,\bu,\cS)$ be the free difference algebra generated by noncommuting variables $\{\bu_n\}_{n\in\bbZ}$, where $\bu=\bu_0$ may be multi-component, with shift automorphism  $\cS(\bu_n)=\bu_{n+1}$. Difference operators are {finite} sums $A=\sum_{n\in\bbZ} a_n\cS^n$, $a_n\in\cA$, and form the ring of Laurent polynomials $\fA=\cA[\cS,\cS^{-1}]$. For $A\in\fA$, let $\pi_+(A)$ denote the projection on its polynomial part in $\cS$, $\phi^{(n)}(A)=a_n$ its $\cS^n$-coefficient, and $\phi(A)$ the set of its nonzero coefficients.

Consider an integrable   hierarchy $\partial_{t_k}(\bu )= \bF^{(k)} $,  admitting a Lax representation
\begin{equation}\label{lax}
\partial_{t_k}L=[A^{(k)},L],\qquad k\in\bbN,
\end{equation}
where $L\in\fA$ is monic, and $A^{(k)}\in \pi_+(\fA)$. It defines derivations $\partial_{t_k}$ of $\cA$, commuting  with $ \cS$.\renewcommand{\theequation}{\arabic{equation}}
Let $\cL=\sum_n \alpha_nL^n$ be a finite sum with $\alpha_n\in\bbC,\, n\in\bbZ$. Negative powers of $L$ are well defined in the ring of formal Laurent series $\cA[\cS][[\cS^{-1}]]$.   For $m\in\bbZ$, define the difference operators
\[
\Lambda_m=\pi_+(\cL\cS^{-m})\cS^m,\qquad \varDelta _m=[\Lambda_m,L],
\]
and the two-sided difference ideal  $\cJ_m=\langle \{\cS^n(\phi(\varDelta_m))\}_{n\in\bbZ}\rangle\subset \cA$.
\begin{thm}\label{thm1}
For all $k\in\bbN,\, m\in\bbZ$, the ideal $\cJ_m\subset\cA $ is $\partial_{t_k}$-stable, and
\begin{equation}\label{lammt}
    \partial_{t_k}(\Lambda_m)-[A^{(k)},\Lambda_m]\in\cJ_m[\cS,\cS^{-1}].
\end{equation}
\end{thm}

Therefore, the derivations $\partial_{t_k}$ descend to the quotient algebra $\cA_{\cJ_m}=\cA/\cJ_m$, reducing the hierarchy to $\cA_{\cJ_m}$. Equations \eqref{lammt} provide Lax representations for the reduced hierarchy and yield its first integrals which are invariants of the integrable map defined by $\phi(\varDelta_m)$.

{\bf 2.} As an illustration, consider the Volterra hierarchy with Lax representation \eqref{lax}, where
\begin{equation}\label{laxVol}
    L=\cS+u\cS^{-1} ,\qquad A^{(k)}=(L^{2k})_+,
\end{equation}
and the first system of the Volterra hierarchy is
\begin{equation}\label{vol1}
  \partial_{t_1}u_n=u_{n+1}u_n-u_nu_{n-1},\qquad n\in\bbZ.
\end{equation}
Let $\cL(N)=\sum_{k=0}^N \alpha_{N-2k}L^{N-2k}$, where $N\in\bbN$ and $\alpha_n\in\bbC$.
The coefficients $\ell(n,m)$  of
$
L^n=\sum_{k\ge0}\ell(n,n-2k)\cS^{n-2k}
$
are uniquely determined by the recurrence relation
\begin{equation}\label{recell}
    \ell(n,m)=\left\{ \begin{array}{ll}
       \ell(n-1,m+1)u_{m+1}+\ell(n-1,m-1)\qquad  &\mbox{if }\ n>0,  \\
       \ell(n+1,m+1)-\ell(n,m+2)u_{m+2}\qquad  &\mbox{if }\ n<0,
    \end{array}\right.
\end{equation}
subject to the boundary conditions
$  \ell(0,m)=\delta_{0,m},\   \ell(n,m)=0$,    if  $m>n $.  It immediately follows from \eqref{recell} that $\ell(n,n)=1$ and $\ell(n,m)=0$ whenever $n+m$ is odd.

\begin{lemma}\label{lem+} Let $L=\cS+u\cS^{-1}$  (\ref{laxVol}). Then
\begin{equation*}\label{lemstat}
    [\pi_+(L^n\cS^{-m})\cS^m,L]=
   (\cS-1) ( \ell(n,m-2))\cS^{m-1}  +
    (\cS-1)(\ell(n,m-1))\cS^{m  } .
\end{equation*}
\end{lemma}
It follows from Lemma~\ref{lem+} that $\cJ_m=\{0\}$ for all $m>N$, and $\cJ_{2k}=\cJ_{2k-1},\, k\in\bbZ$. Thus, we can restrict ourselves with the ideals $ \cJ_{N-2k},\, k\ge 0 $, that  are generated by the polynomials $\{(\cS-1)\cS^n(J_{N-2k})\}_{n\in\bbZ}$, where
\[
J_{N-2k}=\sum\limits_{m=0}^k \alpha_{N-2m}\ell(N-2m,N-2k-2).
\]
For each $k,\,0\le k\le N$, the flows of the Volterra hierarchy admit canonical reduction to an integrable dynamical systems in  $N$ variables. It follows from \eqref{lammt} that the polynomial  $J_{N-2k}$ is a first integral  of the reduced systems. Hence, in the extended algebra $\cA^\bw=\cA[\bw]$, where $w$ is a constant element ($\cS\bw=\bw$ and $\partial_{t_k}\bw=0$),
we define the $\partial_{t_k}$-stable difference ideal $\cI^\bw_{N-2k}=\langle \{\cS^n (J_{N-2k})-\bw\}_{n\in\bbZ}\rangle\subset \cA^{\bw} $. This yields a reduction  of the Volterra hierarchy to the   algebra $\cA^\bw/\cI^\bw_{N-2k}$ with $N-1$ dynamical variables.

The above constructions become explicit in the commutative case upon passing to the field of fractions of $\cA^{\bw}/\cI^\bw_{N-2k}$. In the simplest non-trivial case, $N=3$ and $ \alpha_3=\alpha_{-1}=1$, for each $k,\ 0\le k\le N$,
the Volterra chain \eqref{vol1} reduces to a dynamical system in the variables $u,u_1$, admitting   a  first integral $H_k$ that is invariant under the recursion relations defined by $\{\cS^n (J_{N-2k})=\bw\}_{n\in\bbZ} $:
\begin{equation*}
    \begin{array}{lll}
  k=0:        &\!\!\! \left\{\begin{array}{l}   \partial_{t_1}u =2 u_1 u +u^2 -\bw  u,\\
  \partial_{t_1} u_1= -2u_1u-u_1^2+\bw  u_1;
  \end{array}\right.
       &\!\!\!\begin{array}{l}
       H_0 =u_1  u(u_1+u-w),\\  \cS^n(  u_{ 1}+u +u_{ -1})=\bw . \end{array}\\ \\
 k=1:       &\!\!\! \left\{\begin{array}{l}    \partial_{t_1}u =2 u_1 u +u^2 +\alpha_1 u-\bw ,\\ \partial_{t_1} u_1= -2u_1u-u_1^2-\alpha_1 u_1+\bw;
  \end{array}\right.
       &\!\!\!\begin{array}{l}
       H_1 =(u_1  u-\bw)(u_1+u+\alpha_1),\\
       \cS^n( u_{1} u+u^2 +uu_{-1}+\alpha_1 u)=\bw   . \end{array}   \\ \\

 k=2:       &\!\!\! \left\{\begin{array}{l}
 \partial_{t_1}u =  u_1 u -( u+\bw)u_1^{-1}, \\
 \partial_{t_1} u_1= - u_1u+ ( u_1+\bw)u^{-1};
  \end{array}\right.
       &\!\!\!\begin{array}{l}
        H_2=u_1+u+ u_1^{-1}+ u^{-1} +w u_1^{-1} u^{-1},\\
      \cS^n(  u_{1}uu_{-1}-  u)=\bw  . \end{array} \\ \\

 k=3:       &\!\!\! \left\{\begin{array}{l}
  \partial_{t_1}u = u\dfrac{  \alpha _{-3}(2  u_1+ u) -u_1^2+w}{\alpha _{-3}-u_1} ,\\
 \partial_{t_1} u_1= u_1\dfrac{ \alpha _{-3}( 2   u +  u_1)-u ^2+w}{u-\alpha _{-3} };
  \end{array}\right.
       &\!\!\!\begin{array}{l}
        H_3=\dfrac{\left( u -\alpha _{-3}\right) \left(  u_1-\alpha _{-3}\right)}{u  u_1 \left(\alpha _{-3} u +\alpha _{-3} u_1+w\right)},\\
      \cS^n(u_{1}u_{-1}-\alpha_{-3} (u_{1}+u+u_{-1}))=\bw  . \end{array}
  \end{array}
\end{equation*}
The cases $k=0,1,2$ correspond respectively to: the period-3 reduction, the stationary flow $\partial_{t_2}u_n+\alpha_1\partial_{t_1}u_n=0$, and the commutative Kontsevich system~\cite{wolf}. The case $k=3$ appears to be a new integrable reduction of the Volterra system. Solutions are expressed in elliptic functions, and the associated integrable second-order recurrence maps extend them to the full Volterra chain~\eqref{vol1}.

For $N=2g+1$ and $N=2g+2$, the reductions give systems of $2g$ and $2g+1$ equations, admitting $g$ and $g+1$ first integrals together with integrable maps of orders $2g$ and $2g+1$, respectively.

The approach is applicable to a broad class of Lax-integrable differential-difference hierarchies \eqref{lax}, in both commutative and non-commutative settings.

 \bibliographystyle{unsrt}
\bibliography{short}
%\end{thebibliography}

{\bf A.\,V.~Mikhailov (А.\,В.~Михайлов)}

University of Leeds, Leeds, UK

{\it E-mail}: a.v.mikhailov@leeds.ac.uk

\end{document}

